\documentclass[
journal=jacsat, 
manuscript=article]{achemso}

\usepackage[version=3]{mhchem} 
 \usepackage{bbm}
\usepackage{hyperref}
\usepackage{multirow}
\usepackage{caption}
\usepackage{graphicx}
\usepackage{booktabs}
\usepackage{amsmath}
\usepackage{mathtools}
\usepackage{enumitem}   
\usepackage{makecell}
\usepackage[section]{placeins}
\numberwithin{table}{section}
\numberwithin{figure}{section}
\usepackage{color,soul}
\usepackage{amsthm}
\theoremstyle{definition}
\usepackage{subfig}
\usepackage{float}
\usepackage{amsmath}
\usepackage{titlesec}

\author{Pantelis Loupos}
\email{p-loupos@kellogg.northwestern.edu}
\affiliation
{Kellogg School of Management, Northwestern University, Evanston, IL 60208}
\author{Alexandros Nathan}
\email{anathan@u.northwestern.edu}
\affiliation
{McCormick School of Engineering, Northwestern Universty, Evanston, IL 60208}

\title[\texttt{achemso} demonstration]
{The Structure and Evolution of an Offline Peer-to-Peer Financial Network}

\begin{document}

\begin{abstract}
In this work, we investigate the structure and evolution of a peer-to-peer (P2P) payment application. A unique aspect of the network under consideration is that the edges among nodes represent financial transactions among individuals who shared an offline social interaction. Our dataset comes from Venmo, the most popular P2P mobile payment service. We present a series of static and dynamic measurements that summarize the key aspects of any social network, namely the degree distribution, density and connectivity. We find that the degree distributions do not follow a power-law distribution, confirming previous studies that real-world social networks are rarely scale-free. The giant component of Venmo is eventually composed of 99.9\% of all nodes, and its clustering coefficient reaches 0.2. Last, we examine the “topological” version of the small-world hypothesis, and find that Venmo users are separated by a mean of 5.9 steps and a median of 6 steps.
\end{abstract}

\section{Introduction}

Over the last two decades, there has been a surge of research on complex networks. The availability of high quality network data as a result of the rise of the internet and information technology, has facilitated the computational analysis of social networks. Examples of networks that have been studied include scientific co-authorship networks, online social networks and employee communication networks \cite{clauset2009power,barabasi2002evolution,mislove2007measurement}. While there have been several large-scale studies that focus on the evolution of online social networks \cite{Kumar:2006:SEO:1150402.1150476,Kumar:2004:SEB:1035134.1035162,gong2012evolution}, offline social network data have been scarce and small in scale.

In this work, we study the structural evolution of Venmo, the most popular peer-to-peer (P2P) payment application. Venmo is an Event-Based Social Network (EBSN) \cite{liu2012event}, as it reflects offline face-to-face social interactions in an online social network. In other words, for a user to use Venmo, they need first to share an experience with another user offline, which they will later publish online. Venmo allows peers to send money to each other for a variety of social activities, such as splitting the bill after a group outing at a restaurant or sharing a cab. One of the most distinctive characteristics of Venmo is that users are required to accompany their transactions with a description of what the money was used for. Studies of these description messages\footnote{https://lendedu.com/blog/venmo} have shown that some of the most popular uses of the service include food, transportation and utilities. 

To the best of our knowledge, this is the first time in the social networks literature that we are able to study how financial activity is connected to social activity. By following the money trail in a P2P setting, we can gain insights into the interactions among connections who engage in shared experiences. Further, the nature of the activities Venmo is used for implies that users who transact with each other are often in close geographic proximity, something that is known to have a fundamental effect on social ties: the probability of a new friendship is inversely proportional with the spatial distance between people \cite{brown2012online}.

Our investigation is organized into two parts. In the first part, we present a series of static measurements that characterize the degree distribution of Venmo. Since Venmo is a directed social network, we look at three degree sequences: (a) degree (undirected), (b) in-degree and (c) out-degree. We find that all degree sequences do not follow a power-law distribution, confirming previous studies that real-world social networks are rarely scale-free \cite{broido2018scale}. 

In the second part, we present a series of dynamic measurements that characterize the structural evolution of Venmo's network. These include network density, clustering coefficient, component structure and degrees of separation. Unlike online social networks such as Flickr and Yahoo! 360 \cite{Kumar:2006:SEO:1150402.1150476}, the density of Venmo's network shows a steady increase over time. Furthermore, the giant component is eventually composed of 99.9\% of all nodes, and the clustering coefficient reaches 0.2, which is also common in online social networks. Last, we examine the “topological” version of the small-world hypothesis, and find that Venmo users are separated by a mean of 5.9 steps and a median of 6 steps. A previous study \cite{leskovec2008planetary} of the Microsoft Messenger instant-messaging system found that users were separated by a mean of 6.6 steps and a median of 7 steps. Later on, a study \cite{backstrom2012four} of the Facebook social graph found that the average degrees of separation are 4.5, claiming that online social networks have made the world much smaller. Our finding shows that our world is indeed smaller than we previously believed, but not as much smaller as the Facebook study claimed it to be.

The remainder of this paper is organized as follows: Section~\ref{Data} provides a short overview of our Venmo dataset; in Section \ref{Results} we present our results; finally, Section \ref{Conclusions} concludes with a summary and a discussion of our findings.

\section{Data Overview}\label{Data}

This work uses data from Venmo, a P2P mobile payment service owned by PayPal. Founded in 2009, Venmo operates in the United States and has succeeded in transforming financial transactions into a sharing experience. It allows people to easily transfer money electronically, replacing the traditional wallet of cash and credit cards. What makes Venmo really unique though is its social nature. Upon logging into the application, users gain access to a Facebook-like news feed, which is composed of public transactions. The individual who initiates the transaction is required to accompany the post with a description of what the money was used for, while the dollar amount is left out for privacy reasons. Other users may "like" or comment on the transactions that appear on their news feed. Although the public news feed is entirely open to any Venmo customer, users may opt to hide their transactions by adjusting their privacy settings. However, according to Dan Schulman, CEO of PayPal, "90\% of transactions are shared" \footnote{http://fortune.com/2017/11/17/dan-schulman-paypal-venmo/}. For our analysis, we have collected the complete transaction history of 1,765,776 users comprising of more than 120 million transactions through Venmo's API \cite{kraft2014security}. We are interested in examining the connections between real people, and thus, we exclude from our analysis Venmo accounts that belong to business owners and charity organizations. This leaves us with 1,748,119 accounts that have less than 78 connections, which covers 99\% of the user population. An appealing aspect of our dataset is that we have access to the complete evolution of Venmo's social network, from its inception in 2009 up to late 2016.

\section{Results}\label{Results}

In this section, we present some key measurements that characterize the overall properties of a network. First, we introduce some notation: let $G_t \coloneqq (V_t, E_t)$ be a dynamic social network, where $V_t$ is the set of nodes that have joined the network by time $t$, and $E_t$ consists of all edges that have been formed by time $t$. We treat time as a sequence of discrete monthly intervals. In what follows, we present results for the case where nodes can only be added in the network, that is for $t_2 \geq t_1$, we have that $V_{t_2} \geq V_{t_1}$. 

\subsection{Degree Distribution}

\begin{figure}[H]
\centering
 
\subfloat{
	\label{subfig:correct}
	\includegraphics[width=0.5\textwidth]{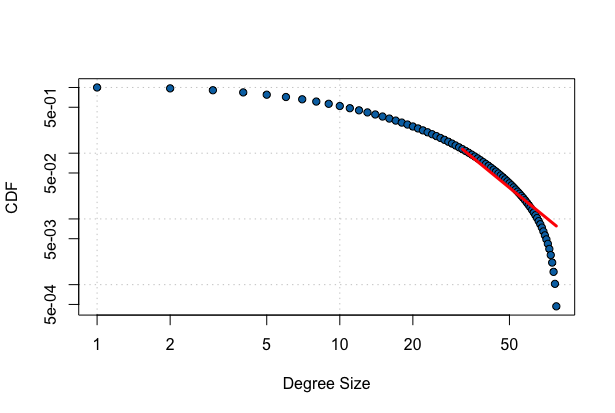} } 

\subfloat{
	\label{subfig:nonkohler}
	\includegraphics[width=0.5\textwidth]{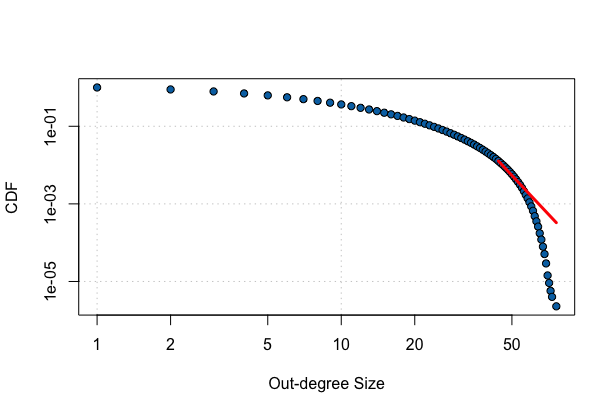}} 
	
\subfloat{
	\label{subfig:notwhitelight}
	\includegraphics[width=0.5\textwidth]{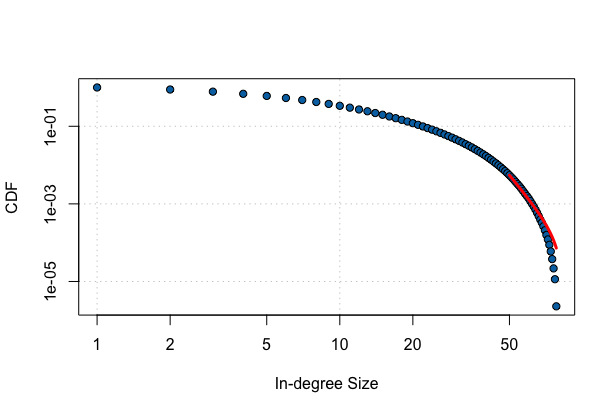} }

\caption{Degree, In-Degree and Out-Degree distributions. We can safely reject the scale-free hypothesis as the p-value is $\approx$0, $\approx$0 and 0.01 for the degree, out-degree and in-degree, respectively.}
\label{fig:degreeDist}
 
\end{figure}

The degree distribution of a network is one of its most important properties. Given that Venmo's transactions are directed in nature, we look at three degree sequences: (a) degree (undirected), (b) in-degree and (c) out-degree. The degree distributions were computed for a static snapshot of the network, when it has reached its steady state (month 74). Figure~\ref{fig:degreeDist} shows the three degree plots. 

Following the approach in Broido et al and Clauset et al. \cite{broido2018scale,clauset2009power}, we fit and test for the plausibility of power laws. We find the p-value to be $\approx$0, $\approx$0 and 0.01 for the degree, out-degree and in-degree, respectively. Therefore, we can safely reject the scale-free hypothesis. 

\newpage

\subsection{Density and Densification}

\begin{figure}[H]
\begin{center}
\includegraphics[width=0.5\columnwidth]{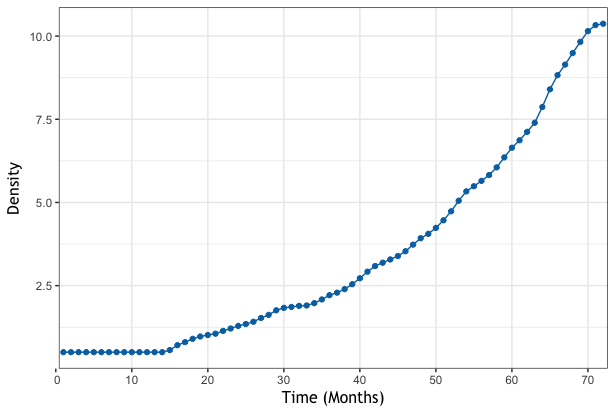}
\caption{Density over time.} 
\label{fig:density}
\end{center}
\end{figure}

We look first at the evolution of density over time. Density is defined as the ratio of undirected edges to nodes. As can be seen in Figure~\ref{fig:density}, density is monotonically increasing as a function of time.  

\begin{figure}[H]
\begin{center}
\includegraphics[width=0.5\columnwidth]{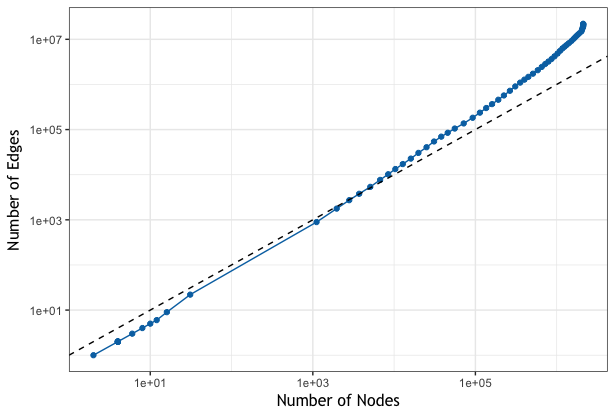}
\caption{DPL plot. The dashed line is the diagonal with slope 1. The best fitted line to our data has a slope of $\alpha = 1.19$} 
\label{fig: DPL}
\end{center}
\end{figure}

Densification was first described in Leskovec et al. \cite{leskovec2005graphs}. Essentially in a network that exhibits densification over time, the average degree also increases. An easy way to show this is via the Densification Power Law plot (DPL) - see Figure~\ref{fig: DPL}. The DPL plot refers to the log-log plot of the number of nodes $V_t$ against the number of edges $E_t$ at several snapshots $t$. We subsequently calculate the slope of the line, $\alpha$, that best fits our data. A value of $\alpha>1$, implies that the average degree increases over time, and in our case it comes out to be $\alpha=1.19$. 

\subsection{Clustering Coefficient}

The clustering coefficient measures the extent to which an individual's friends know each other. As expected, a high clustering coefficient implies a large proportion of triangles (triads) in the network. While there are a few different definitions for the clustering coefficient, we follow the one outlined in Watts et al. \cite{watts1998collective}. Let $a_{ij}$ be the element of adjacency matrix indicating the existence or absence of an (undirected) edge between nodes $i$ and $j$, and $k_{i}$ denote the degree of node $i$. The average clustering coefficient is defined as:

$$C^{avg}={\displaystyle \frac{1}{n}\sum_{i}\sum_{j,k}\frac{a_{ij}a_{jk}a_{ki}}{k_{i}(k_{i}-1)/2}}$$
 
\begin{figure}[H]
\begin{center}
\includegraphics[width=0.5\columnwidth]{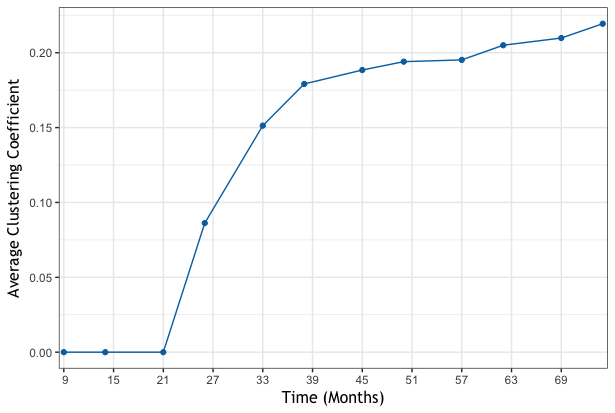}
\caption{Clustering coefficient averaged across all nodes - it's increasing over time.} 
\label{fig:clusteringCoeff}
\end{center}
\end{figure}

Figure~\ref{fig:clusteringCoeff} shows the evolution of the clustering coefficient over time. We see two distinct phases: a sharp increase, followed by a plateau around 0.2. 

\subsection{Component Structure}

\begin{figure}[H]
\centering
 
\subfloat{
	\label{subfig:giant}
	\includegraphics[width=0.5\textwidth]{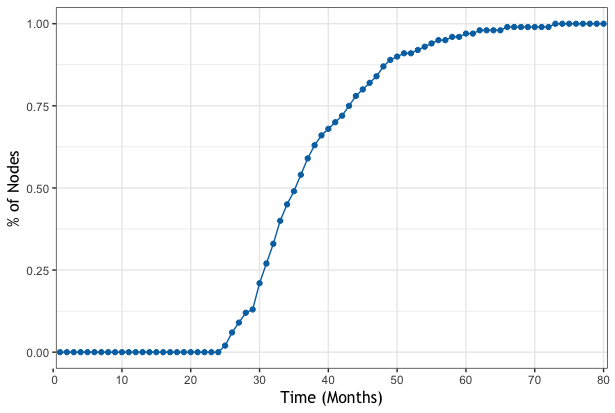} } 
 
\subfloat{
	\label{subfig:Component_evolution}
	\includegraphics[width=0.5\textwidth]{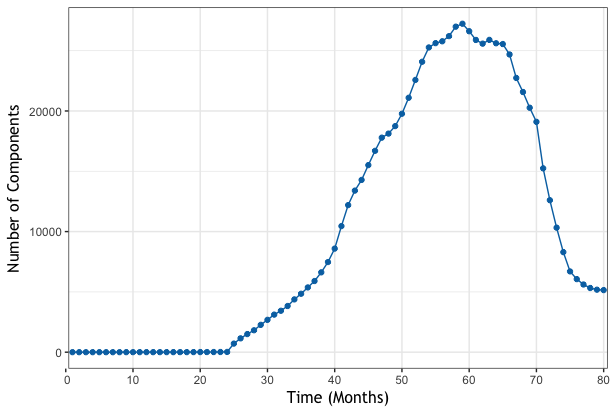} } 
 
\caption{Sub-figure (a) shows fraction of the users that belong to the giant component over time. Sub-figure (b) shows the overall number of components over time.}
\label{fig:components}
 
\end{figure}

\begin{figure}[H]
\begin{center}
\includegraphics[width=0.5\columnwidth]{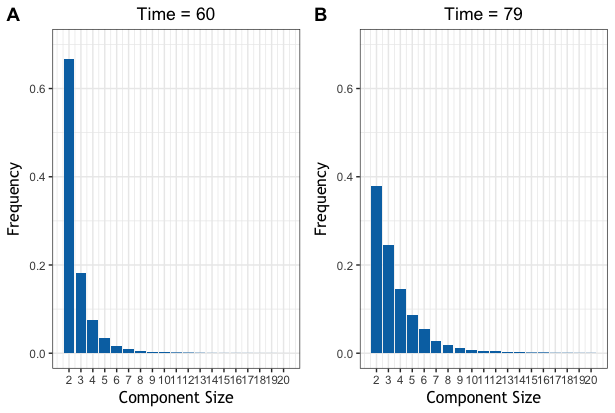}
\caption{Sub-figure A shows the component size distribution at time 60, and sub-figure B shows the same plot at time 79.} 
\label{fig:componentDistSize}
\end{center}
\end{figure}

We investigate here the existence of a giant component, a common feature of most networks \cite{easley2010networks}, as well as the migration patterns of the smallest components. As illustrated in Figure~\ref{subfig:giant}, Venmo's transaction network exhibits a giant component that eventually contains over $99\%$ of all nodes of the network. Figure~\ref{subfig:Component_evolution} shows how the number of components changes over time. It is interesting that the number of components keep increasing over time, only to decrease sharply towards the end - the tipping point is around month 60. Note also that the distribution of the component size undergoes a significant transition following month 60. Figure~\ref{fig:componentDistSize} shows the before and after distributions. While the majority of components at month 60 are of size 2, at month 79 only the larger components survive, with the distribution exhibiting a heavier tail. 

\subsection{Degrees of Separation}

\begin{figure}[h!]
\begin{center}
\includegraphics[width=0.6\columnwidth]{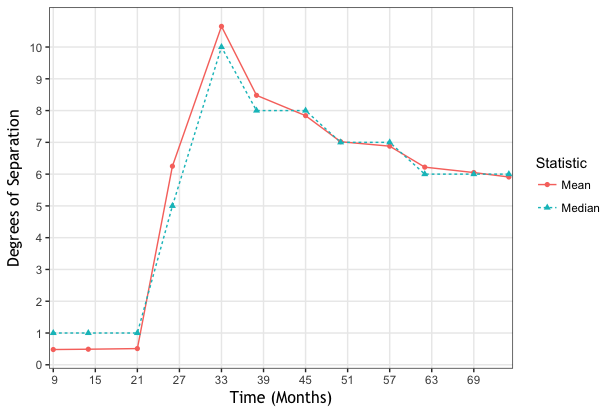}
\caption{Mean and Median Shortest Path Distance over time.} 
\label{fig:ShortestPath}
\end{center}
\end{figure}

Last, we turn our attention in studying the “topological” version of the small-world hypothesis that has long attracted the interest of social scientists. To run our analysis, we only focus on the giant component and use R's Igraph package \cite{igraph} to calculate all shortest paths between two randomly chosen nodes. We repeat this for 2000 random pairs and average the results (we experimented with different values and found that after 2000 our results do not change). Since this calculation is very computationally expensive, we calculate it every six months and not every month as before. As can be seen in Figure~\ref{fig:ShortestPath}, average distance follows three distinct phases. Initially, we have a plateau until month 21. Then, we have a sharp increase, reaching its maximum value at month 33 - the average distance is equal to 10.35 while the median is 10. In the final stage, average distance starts decreasing until it converges to its final value of 5.90 with a median of 6. 

\section{Conclusions and Discussion}\label{Conclusions}

In this work, we investigated the evolution of the structural properties of Venmo. To the best of our knowledge, Venmo is the largest dataset of P2P financial activity ever to be analyzed. Its unique aspect of reflecting offline shared social activities among individuals make its properties documentation worth having. 

Our first finding is that Venmo's degree do not follow a power law. Various studies have documented that power law distributions are common across many networks, such as citation networks \cite{redner1998popular}, the Internet and the Web \cite{faloutsos1999power,kleinberg1999web,broder2000graph,barabasi1999emergence,barabasi2000scale,huberman1999internet,kumar1999trawling}, online social networks \cite{mislove2007measurement} and phone call graphs \cite{abello1998functional}. However, a recent study \cite{broido2018scale} that investigates a 1000 networks found that scale-free real world networks are rare. We add to this observation that real-world networks exhibit a much richer structure.

In the second part, we present a series of dynamic measurements that characterize the structural evolution of Venmo's network. These include network density, clustering coefficient, component structure and degrees of separation. 

Venmo's density shows a steady increase over time. This stands in direct contrast with the findings of Kumar et al. \cite{Kumar:2006:SEO:1150402.1150476}, who observe three different stages: an initial upward trend, followed by a dip, and finally a gradual steady increase. We believe that unlike Yahoo!360 and Flickr, Venmo has continued to grow at an incredible pace ever since 2010, continuously obtaining new users, while at the same time retaining the vast majority of the existing ones. Regarding its densification, the best fitted line has a slope of 1.19. This is consistent with the four networks that Leskovec et al examined \cite{leskovec2005graphs}, and more specific, our value is very close to the value of 1.15 corresponding to affiliation graphs of co-authors in the arXiv.

Venmo's clustering coefficient undergoes two distinct phases: a sharp increase, followed by a plateau around 0.2. This is in contrast with other networks, such as Google+, which shows three phases (decrease, increase and decrease again) \cite{gong2012evolution}.

Furthermore, the giant component is eventually composed of 99.9\% of all nodes. The number of components keep increasing over time, only to decrease sharply towards the end - the tipping point is around month 60. The distribution of the component size undergoes a significant transition following month 60.  While the majority of components at month 60 are of size 2, at month 79 only the larger components survive, with the distribution exhibiting a heavier tail. It is noteworthy that this behavior is consistent with the random graph model introduced by Erdos and Renyi \cite{erdos1960evolution}.

Last, we find that Venmo users are separated by a mean of 5.9 steps and a median of 6 steps. Travers and Milgram in their monumental experiments \cite{milgram1967small,travers1977experimental}, claimed that the degrees of separation across people are six. We should point out here, however, that their results correspond to the "algorithmic" version of the small-world hypothesis, which provides an upper bound on the average distance. Goel et al. \cite{goel2009social} corrected for this bias and found the median shortest path to be 7. A recent study by Leskovec et al. \cite{leskovec2008planetary} investigated the Microsoft Messenger instant-messaging system, a communication graph of 180 million nodes and 1.3 billion edges. They found that users were separated by a mean of 6.6 steps and a median of 7 steps. Later on, Backstrom et al. \cite{backstrom2012four} studied the Facebook social graph and found the average degrees of separation to be 4.5, claiming that the world is even smaller than we expected. Our results come to shed light to all these previous studies. On one hand, we see that social networks tend to deflate the degrees by which people are separated. On the other hand, we find that the world is indeed smaller that we previously believed. In fact, it is 6 degrees separated.

\bibliography{sample}

\end{document}